\documentstyle[preprint,aps,epsf]{revtex}
\tightenlines
\begin{document}
%
%
%
%
\draft
\title{
%
%
\begin{flushright}
\normalsize
KEK-CP-082          \\
KEK Preprint 98-217 \\
January 1999        \\
\end{flushright}
%
%
Non-perturbative determination of quark masses
in quenched lattice QCD
with the Kogut-Susskind fermion action
}
%
%
\author{
      S.~Aoki,$^{\rm 1    }$
  M.~Fukugita,$^{\rm 2    }$
 S.~Hashimoto,$^{\rm 3    }$
K-I.~Ishikawa,$^{\rm 4    }$
  N.~Ishizuka,$^{\rm 1, 5 }$
   Y.~Iwasaki,$^{\rm 1, 5 }$
    K.~Kanaya,$^{\rm 1, 5 }$
    T.~Kaneda,$^{\rm 1    }$
      S.~Kaya,$^{\rm 3    }$
 Y.~Kuramashi,$^{\rm 3    }$
     M.~Okawa,$^{\rm 3    }$
     T.~Onogi,$^{\rm 4    }$
  S.~Tominaga,$^{\rm 3    }$
   N.~Tsutsui,$^{\rm 4    }$
     A.~Ukawa,$^{\rm 1, 5 }$
    N.~Yamada,$^{\rm 4    }$
T.~Yoshi\'{e},$^{\rm 1, 5 }$ \\
(JLQCD Collaboration)
}
\address{
${}^{\rm 1}$ Institute of Physics, University of Tsukuba, Tsukuba, Ibaraki 305-8571, Japan \\
${}^{\rm 2}$ Institute for Cosmic Ray Research, University of Tokyo, Tanashi, Tokyo 188-8502, Japan \\
${}^{\rm 3}$ High Energy Accelerator Research Organization (KEK), Tsukuba, Ibaraki 305-0801, Japan \\
${}^{\rm 4}$ Department of Physics, Hiroshima University, Higashi-Hiroshima, Hiroshima 739-8526, Japan \\
${}^{\rm 5}$ Center for Computational Physics, University of Tsukuba, Tsukuba, Ibaraki 305-8577, Japan \\
}
\date{\today}
%
\maketitle
%
%
\begin{abstract}
We report results of quark masses in quenched lattice QCD with the 
Kogut-Susskind fermion action, employing the Reguralization Independent scheme (RI)
of Martinelli {\it et al.} 
to non-perturbatively evaluate the renormalization factor 
relating the bare quark mass on the lattice to that in the continuum. 
Calculations are carried out at $\beta=6.0$, $6.2$, and $6.4$, from which
we find 
$m^{\overline{\rm MS}}_{ud} (2 {\rm GeV})= 4.23(29) {\rm MeV}$ 
for the average up and down quark mass
and, with the $\phi$ meson mass as input, 
$m^{\overline{\rm MS}}_{s } (2 {\rm GeV})=  129(12) {\rm MeV}$ 
for the strange mass in the continuum limit.
These values are about 20\% larger than
those obtained with the one-loop perturbative renormalization factor.
\end{abstract}
%
%
%
\narrowtext
%
%
%
The values of quark masses are fundamental parameters
of the Standard Model which are not directly accessible 
through experimental measurements. 
Lattice QCD allows their determination through 
a calculation of the functional relation between quark masses and hadron masses.
For this reason a number of lattice QCD calculations have been carried out 
to evaluate quark masses, employing the Wilson, clover 
or Kogut-Susskind (KS) fermion action~\cite{review}.  

An important ingredient in these calculations is the renormalization factor 
relating the bare lattice quark mass to that in the continuum.  
While perturbation theory is often used to evaluate this factor,
uncertainties due to higher order terms 
are quite significant in the range of the QCD coupling constant 
accessible in today's numerical simulations.  
A non-perturbative determination of the renormalization factor is therefore
necessary for a reliable calculation of quark masses, and effort in this 
direction has recently been pursued for the Wilson
and clover fermion actions~\cite{martinelli,jansen,gimenez,becirevic}. 

The need for a non-perturbative determination of the renormalization factor 
is even more urgent for the KS action
since the one-loop correction~\cite{perturbation,KS} is as large as 50\% 
in present simulations.  
In this article we report a study to meet this need~\cite{ishizuka98,kilcup97} :
we calculate the renormalization factor of bi-linear quark operators for the
KS action non-perturbatively using the Reguralization Independent scheme (RI) 
of Ref.~\cite{martinelli} developed for the Wilson/clover actions.
The results for the scalar operator, combined with our previous calculation
of bare quark masses~\cite{JLQCD_M},
lead to a non-perturbative determination of the quark masses in the continuum limit.  

In the RI sheme, the renormalization factor of a bi-linear operator 
${\cal O}$ is obtained from the amputated Green function,
\begin{equation}
  \Gamma_{\cal O}(p)
= S(p)^{-1} \langle 0 | \phi ( p )\ {\cal O}\ \bar{\phi} (p) | 0 \rangle
  S(p)^{-1}
\end{equation}
where the quark two-point function is defined by
$S(p)=\langle 0 | \phi ( p ) \bar{\phi} (p) | 0 \rangle$.
The quark field $\phi (p)$ with momentum $p$ is related 
to the one-component KS field $\chi(x)$ by
$\phi_A (p) = \sum_y {\rm exp}( - i p\cdot y) \chi ( y + aA )$,
where $y_\mu=2a n_\mu$, $p_\mu = 2\pi n_\mu / (aL)$ with 
$-L/4\leq n_\mu<L/4$ and $A_\mu=0,1$.

Bi-linear operators have a form
\begin{equation}
   {\cal O} 
= \sum_{yABab} \bar{\phi}_A^a (y) \overline{( \gamma_S \otimes \xi_F )}_{AB}
            \  U_{AB}^{ab} (y) \phi_B^b (y)
\label{opop}
\end{equation}
where $\overline{( \gamma_S \otimes \xi_F )}$ refers to
Dirac ($\gamma_S$) and KS flavor ($\xi_F$) structure~\cite{KS}, 
and the indices $a$ and $b$ refer to color. 
The factor $U_{AB}^{ab} (y)$ is the product of gauge link variables
along a minimum path from $y+aA$ to $y+aB$.
We note that $U_{AB}(y)$ is absent for scalar and pseudo scalar operators
as these operators are local.

The renormalization condition imposed on $\Gamma_{\cal O}(p)$ is given by
\begin{equation}
   Z_{\cal O}^{\rm RI} (p) \cdot Z_{\phi}(p) 
= {\rm Tr}[ P_{\cal O}^\dagger \Gamma_{\cal O}(p) ]
\label{eq:RI}
\end{equation}
where 
$P_{\cal O}^\dagger=\overline{(\gamma_S^\dagger \otimes \xi_F^\dagger )}$ 
is the projector onto the tree-level amputated Green function.
The wave function renormalization factor $Z_{\phi}(p)$  
can be calculated by the condition $Z_V(p)=1$ for the conserved vector current
corresponding to $\overline{(\gamma_\mu\otimes I)}$.
Since the RI scheme explicitly uses the quarks in external states, 
gauge fixing is necessary. 
We employ the Landau gauge throughout the present work.

The relation between the bare operator on the lattice
and the renormalized operator in the continuum takes the form,
\begin{equation}
     {\cal O}_{\overline{\rm MS}}(\mu)
= U_{\overline{\rm MS}}(\mu,p) Z^{\overline{\rm MS}}_{\rm RI} (p) 
        / Z_{\cal O}^{\rm RI}(p)
    {\cal O}
\label{Lat_Cont_RI}
\end{equation}
where $U_{\overline{\rm MS}}(\mu,p)$ is the renormalization-group running 
factor in the continuum from momentum scale $p$ to $\mu$.
We adopt the naive dimensional regularization (NDR) 
with the modified minimum subtraction scheme ($\overline{\rm MS}$) in the continuum.
The factor $Z^{\overline{\rm MS}}_{\rm RI}(p)$ provides matching
from the ${\rm RI}$ scheme to the $\overline{\rm MS}$ scheme. 
These two factors are calculated perturbatively in the continuum.
For our calculation of the quark mass we apply the relation (\ref{Lat_Cont_RI})
in the scalar channel in the chiral limit, 
{\it i.e}, $1/Z_m^{\rm RI} = Z_{S}^{\rm RI}$.

Our calculations are carried out in quenched QCD.
Gauge configurations are generated with the standard plaquette action  
at $\beta=6.0$, $6.2$, and $6.4$ on an $32^4$ lattice.
For each $\beta$ we choose three bare quark masses tabulated 
in Table~\ref{ParamT} where the inverse lattice spacing $1/a$ is taken 
from our previous work~\cite{JLQCD_M}.

We calculate Green function for $15$ momentum 
in the range $0.038553 \leq (ap)^2 \leq 1.9277$.
Quark propagators are evaluated with a source in momentum eigenstate. 
We find that the use of such a source results in very small statistical 
errors of $O(0.1\%)$ in Green functions.

The RI method completely avoids
the use of lattice perturbation theory. 
We do not have to introduce any ambiguous scale, such as $q^*$~\cite{lepage},
to improve on one-loop results. 
An important practical issue, however, is whether 
the renormalization factor can be extracted from a momentum range
$\Lambda_{\rm QCD} \ll p \ll O(1/a)$
keeping under control the higher order effects in 
continuum perturbation theory, 
non-perturbative hadronization effects, and the discretization error on the lattice.
These effects appear as $p$ dependence of the renormalization factor 
in (\ref{Lat_Cont_RI}), which should be absent 
if these effects are negligible.

\begin{figure}
\centerline{\epsfxsize=14.0cm \epsfbox{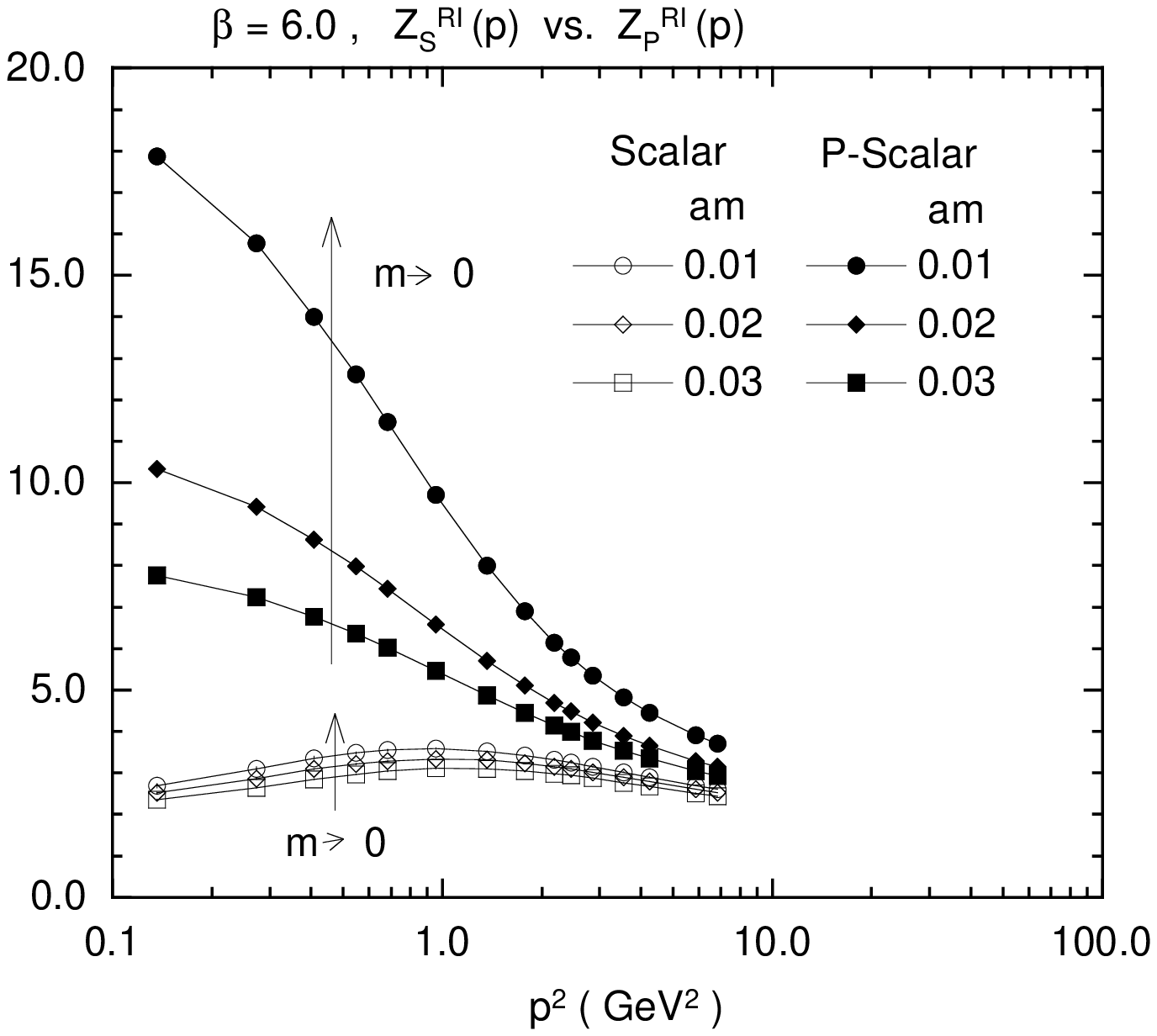}}
%
\caption{\label{ZsZp_D}
The renormalization factor for the scalar $Z_S^{\rm RI}(p)$ 
and the pseudo scalar $Z_P^{\rm RI}(p)$.
}
\end{figure}

In Fig.~\ref{ZsZp_D} we compare 
the scalar renormalization factor $Z_S^{\rm RI} (p)$
with that for pseudo scalar $Z_P^{\rm RI}(p)$ for three values
of bare quark mass $a m$ at $\beta=6.0$.
From chiral symmetry of the KS fermion action,
we naively expect a relation $Z_S^{\rm RI}(p)= Z_P^{\rm RI}(p)$ 
for all momenta $p$ in the chiral limit.
Clearly this does not hold with our result toward small momenta,
where $Z_P^{\rm RI}(p)$ rapidly increases as $m\to 0$,
while $Z_S^{\rm RI}(p)$ does not show such a trend.

To understand this result, we note that chiral symmetry of KS fermion leads 
to the following identities between the amputated Green function
of the scalar $\Gamma_S (p)$, pseudo scalar $\Gamma_P (p)$,
and the quark two-point function $S(p)^{-1}$ : 
\begin{eqnarray}
&& \Gamma_S (p) = \frac{ \partial }{ \partial m } S(p)^{-1}  \\
&& \Gamma_P (p) = \frac{1}{2m}
   \bigl[ 
                  \overline{(\gamma_5 \otimes \xi_5 )} S(p)^{-1} 
     +  S(p)^{-1} \overline{(\gamma_5 \otimes \xi_5 )}
   \bigr]
\label{IDIDID}
\end{eqnarray}
We also find numerically that the quark two-point function can be well 
represented by 
\begin{equation}
   S(p)^{-1} 
   \sim \sum_\mu \overline{( \gamma_\mu \otimes I )} \Sigma_\mu^\dagger (p) i C_\mu (p) 
+ M(p)
\label{SPSPSP}
\end{equation}
with two real functions $C_\mu (p)$ and $M(p)$,
where 
$\Sigma_\mu^\dagger (p)
= \cos ( a p_\mu ) - i \overline{( \gamma_\mu \gamma_5 \otimes \xi_\mu \xi_5 )} \sin (ap_\mu)$.
From (\ref{IDIDID}), (\ref{SPSPSP}), and (\ref{eq:RI})
we obtain the relations between the renormalization factors and $M(p)$,
\begin{eqnarray}
 &&  Z_S^{\rm RI}(p) \cdot Z_\phi(p) = { \partial M(p) } / { \partial m }   \\
 &&  Z_P^{\rm RI}(p) \cdot Z_\phi(p) = {  M(p) } / {  m }
\label{ZP}
\end{eqnarray}

\begin{figure}
\centerline{\epsfxsize=14.0cm \epsfbox{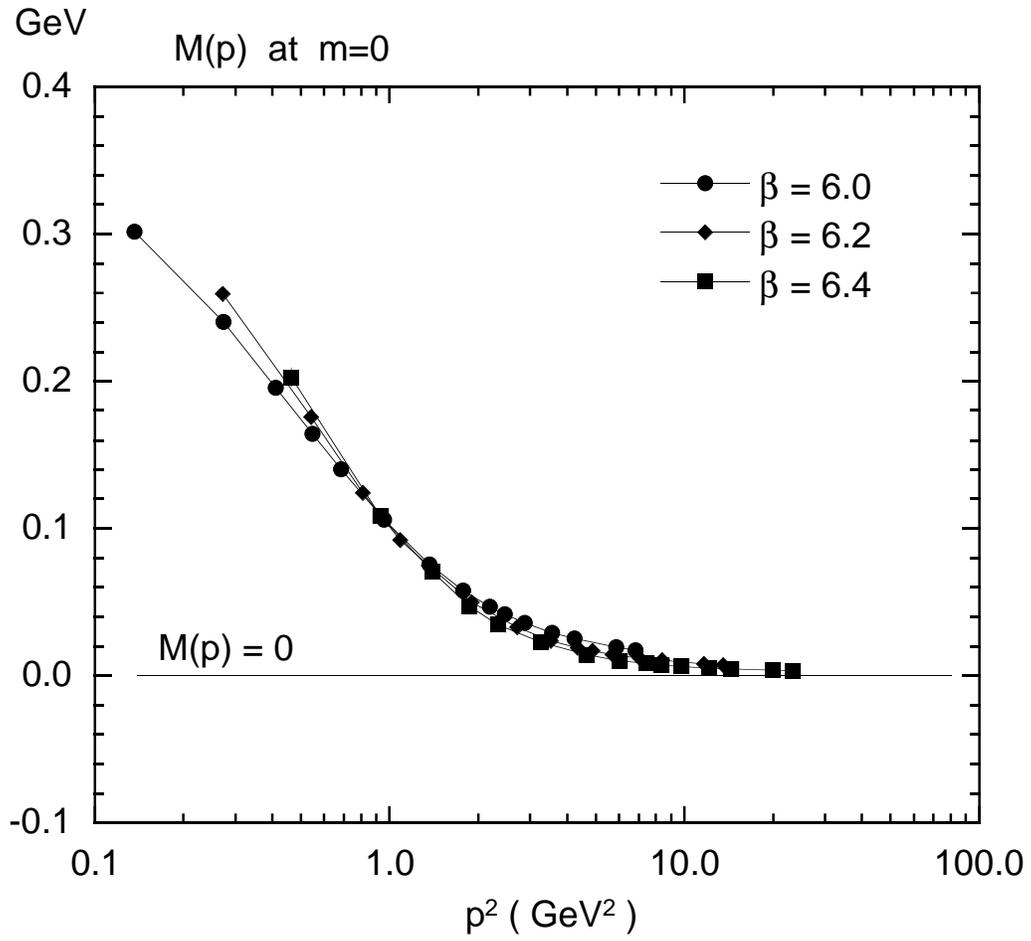}}
%
\caption{\label{Mp}
$M(p)$ in the chiral limit.
}
\end{figure}

In Fig.~\ref{Mp} $M(p)$ in the chiral limit obtained by a linear extrapolation in $m$ is plotted.
It rapidly dumps for large momenta, but largely increases toward small momenta.
Combined with (\ref{ZP}) this implies that $Z_P^{\rm RI}(p)$
diverges in the chiral limit for small momenta, which is consistent with
the result in Fig.~\ref{ZsZp_D}.

The function $M(p)$ is related to the chiral condensate as follows :
\begin{equation}
   \langle \phi \bar{\phi} \rangle = \sum_p {\rm Tr}[ S(p) ]
    = \sum_p \frac{ M(p) }{ \sum_\mu C_\mu(p)^2 + M(p)^2 }
\end{equation}
A non-vanishing value of $M(p)$ for small momenta would lead to
a non-zero value of the condensate.
Therefore the divergence of $Z_P^{\rm RI}(p)$ near the chiral limit
is a manifestation of spontaneous breakdown of chiral symmetry;
it is a non-perturbative hadronization effect
arising from the presence of massless Nambu-Goldstone boson
in the pseudo scalar channel.

While we do not expect the pseudo scalar meson to affect 
the scalar renormalization factor $Z_S^{\rm RI}(p)$, 
as indeed observed in the small quark mass dependence seen 
in Fig.~\ref{ZsZp_D}, 
the above result raises a warning that $Z_S^{\rm RI}(p)$ may still be 
contaminated by hadronization effects for small momenta. 

\begin{figure}
\centerline{\epsfxsize=14.0cm \epsfbox{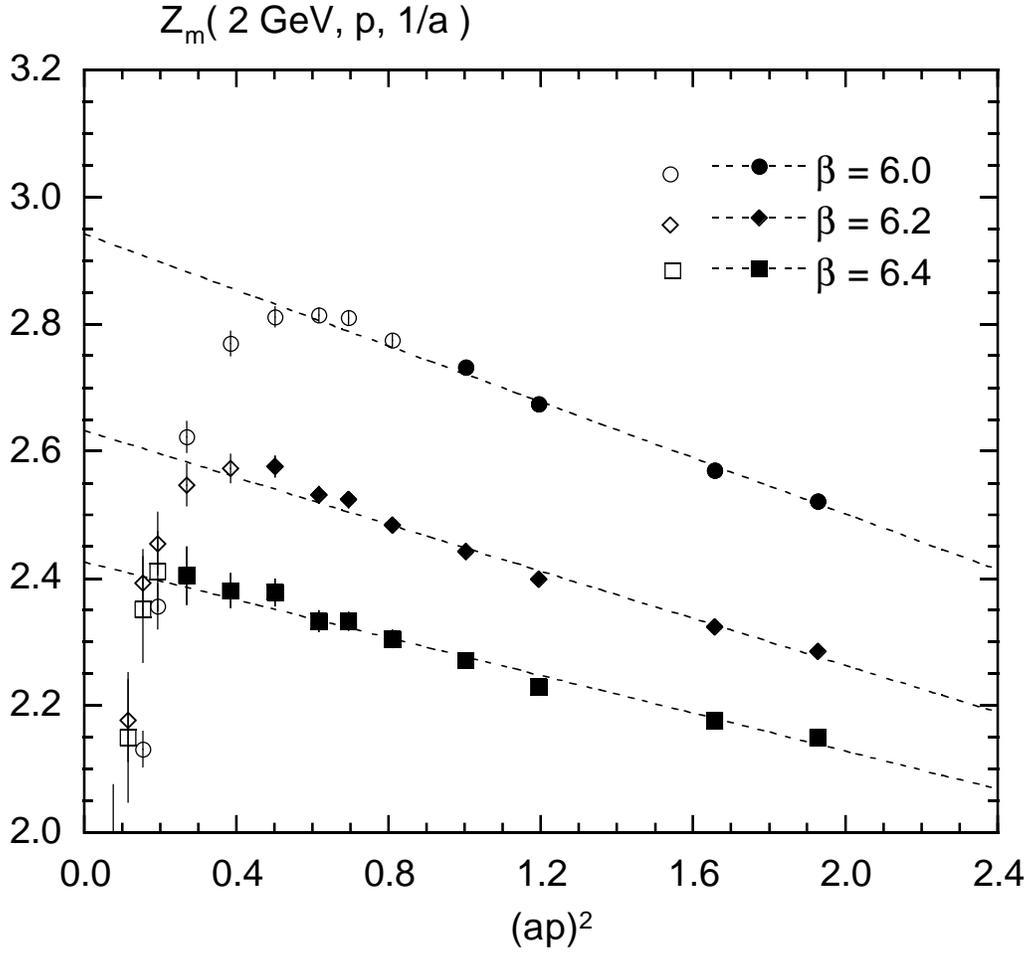}}
%
\caption{\label{Zmq}
The ratio $Z_m(\mu,p,1/a)$ at $\mu=2{\rm GeV}$.
For each $\beta$ the filled data points are used
for linear interpolation in $(ap)^2$.
}
\end{figure}

In Fig.~\ref{Zmq} we show the momentum dependence of 
$Z_m( \mu, p, 1/a ) \equiv
U_{\overline{\rm MS}}(\mu,p) Z^{\overline{\rm MS}}_{\rm RI}(p)
Z_S^{\rm RI}(p)$
which is the renormalization factor from the bare quark mass on the lattice
to the renormalized quark mass at scale $\mu$ in the continuum.
Here we set $\mu=2{\rm GeV}$ and use the three-loop formula~\cite{three_loop}
for $U_{\overline{\rm MS}}$ and $Z^{\overline{\rm MS}}_{\rm RI}$.
While $Z_m( \mu, p, 1/a)$ should be independent of the quark momentum $p$,
our results show a sizable momentum dependence which is almost linear 
in $(ap)^2$ for large momenta (filled symbols in Fig.~\ref{Zmq}).  

For small momenta we consider that the momentum dependence arises from 
non-perturbative hadronization effects on the lattice and the higher order effects 
in continuum perturbation theory. 
It is very difficult to remove these effects from our results.

Toward large momenta, however, these effects are expected to disappear. 
The linear dependence on $(ap)^2$, which still remains,
should arise from the discretization error on the lattice, {\it i.e.},
\begin{equation}
  Z_m( \mu, p, 1/a )
= m^{\overline{\rm MS}}(\mu) / m + (ap)^2 Z_{\rm H} + O(a^4)
\label{p2-dep-rem}
\end{equation}
with the constant $Z_{\rm H}$ corresponding to the mixing
to dimension 5 operators on the lattice.

This relation implies that, if we take a continuum extrapolation of 
$Z_m(\mu, p, 1/a ) m$ at a fixed physical momentum $p$, the discretization 
error in $Z_m$ is removed.
This procedure also removes the $a^2$ discretization error
in the lattice bare quark mass $m$ itself reflecting that in hadron masses.

The momentum $p$ should be chosen in the region where the linear dependence on 
$(ap)^2$ is confirmed in our results.
This region starts from a similar value of $p^2\approx 3 {\rm GeV}^2$ for 
the three $\beta$ values, 
and extends to $p^2 \approx 1.9 /a^2$, the highest momentum measured. 
Hence we are able to use only a rather 
narrow range $3 {\rm GeV}^2 < p^2 < 6.6 {\rm GeV}^2$,
the upper bound dictated by the value of $1.9/a^2$ for the largest lattice spacing at
$\beta=6.0$.

\begin{figure}
\centerline{\epsfxsize=14.0cm \epsfbox{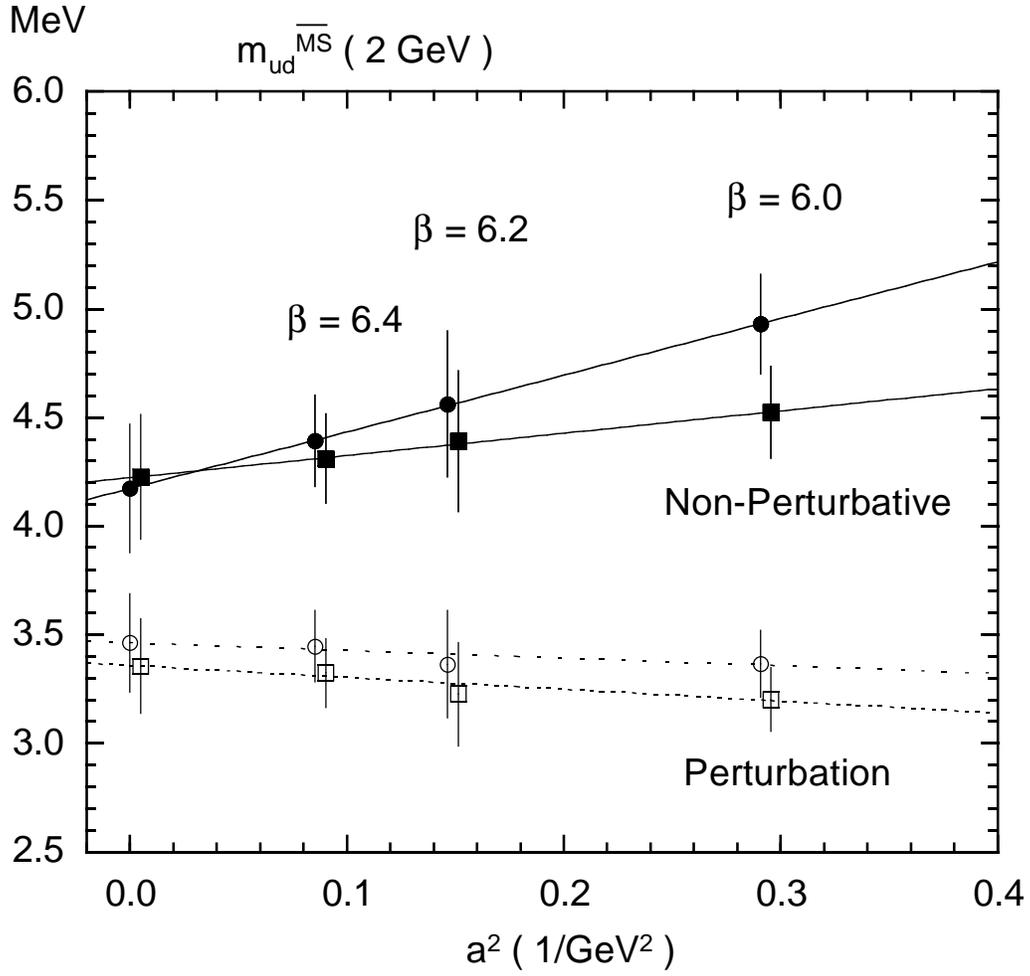}}
%
\caption{\label{Mq}
The final results of the light quark mass $m^{\overline{\rm MS}}_{ud}(2{\rm GeV})$ .
}
\end{figure}

In Fig.~\ref{Mq} we show the continuum extrapolation 
for the averaged up and down quark mass at $\mu=2{\rm GeV}$.  
Filled circles are obtained for $p=1.8{\rm GeV}$
and squares for $p=2.6{\rm GeV}$ for which the value of $Z_m$
is obtained by a linear fit in $(ap)^2$
employing the filled points in Fig.~\ref{Zmq}. 
The bare quark mass~\cite{JLQCD_M} is determined
by a linear extrapolation of pseudo scalar meson mass squared 
in the Nambu-Goldstone channel $\overline{(\gamma_5\otimes \xi_5)}$ 
and that of vector meson mass in the $VT$ channel 
$\overline{(\gamma_i\otimes \xi_i)}$ 
to the physical point of $\pi$ and $\rho$ meson masses. 

We observe that the continuum extrapolation completely removes 
the momentum dependence of the quark mass at finite lattice spacings. 
Furthermore the values are substantially larger than 
those obtained with one-loop perturbation theory 
(open circles for $q^*=1/a$ and squares for $q^*=\pi/a$ ).
Making a linear extrapolation in $a^2$,
our final result in the continuum limit is 
\begin{equation}
  m^{\overline{\rm MS}}_{ud} (2 {\rm GeV}) = 4.23(29) {\rm MeV}.
\end{equation}
where we adopt the value for $p=2.6{\rm GeV}$ since this is the largest 
momentum accessible and the momentum dependence is negligible.
This value is about $20\%$ larger than the perturbative estimates :
$3.46(23){\rm MeV}$ for $q^*=1/a$ and $3.36(22){\rm MeV}$ for $q^*=\pi/a$.
We collect the values of renormalization factor and quark masses in 
Table~\ref{Table_Zm} and \ref{Pesult_mud}. 

Applying our renormalization factor to the strange quark mass,
we obtain 
\begin{eqnarray}
 m^{\overline{\rm MS}}_{s} (2 {\rm GeV}) =& 106.0(7.1) {\rm MeV} \quad & \mbox{ for $m_K$    }  \\
                                         =& 129(12)    {\rm MeV} \quad & \mbox{ for $m_\phi$ } 
\end{eqnarray}
where we use $K$ or $\phi$ meson mass to determine the bare strange mass.
Results from perturbative estimation are given in Table~\ref{Pesult_ms}.

The CP-PACS Collaboration recently reported the results~\cite{CP-PACS} 
$m^{\overline{\rm MS}}_{ud} (2 {\rm GeV}) = 4.6(2) {\rm MeV}$, 
$m^{\overline{\rm MS}}_{s } (2 {\rm GeV}) = 115(2) {\rm MeV}(m_K)$ and 
$143(6) {\rm MeV} (m_\phi)$ from a large-scale precision simulation of hadron 
masses with the Wilson action. 
Our values are 10\% smaller, which may be due to the use of one-loop 
perturbative renormalization factor in the CP-PACS analysis. 

This work is supported by the Supercomputer Project No.32 (FY1998)
of High Energy Accelerator Research Organization (KEK),
and also in part by the Grants-in-Aid of the Ministry of Education 
(Nos.~08640404, 09304029, 10640246, 10640248, 10740107, 10740125).
S.K. and S.T. are supported by the JSPS Research Fellowship.
%
%

%
%
%
%
%
%
%
%
%
%
\begin{table}
\begin{center}
\begin{tabular}{llllll}
$\beta$  &  $1/a ( {\rm GeV} )$   &   \multicolumn{3}{c}{ $a m$ } & $\#$ conf. \\
\hline
6.0      & 1.855(38)              &   0.010  &  0.020  & 0.030       & 30 \\
6.2      & 2.613(87)              &   0.008  &  0.015  & 0.023       & 30 \\
6.4      & 3.425(72)              &   0.005  &  0.010  & 0.020       & 50 \\
\end{tabular}
\end{center}
\caption{ \label{ParamT}
Run parameters.
}
\end{table}
%
%
\begin{table}
\begin{center}
\begin{tabular}{lllll}
$\beta$  &  \multicolumn{2}{c}{NP.}                &  \multicolumn{2}{c}{P.}  \\
         &  $p=1.8 {\rm GeV}$ & $p=2.6 {\rm GeV}$  & $q^*=1/a$   &  $q^*=\pi/a$ \\
\hline	          
6.0      &  2.735(16)  &  2.509(18)  &  1.867 & 1.776 \\
6.2      &  2.5451(91) &  2.449(10)  &  1.877 & 1.800 \\
6.4      &  2.385(12)  &  2.339(12)  &  1.871 & 1.804 \\
\end{tabular} 
\end{center}
\caption{ \label{Table_Zm}
The non-perturbative (NP) and the perturbative (P) renormalization factors
of the quark mass $Z_m(\mu,p,1/a)$ at $\mu=2{\rm GeV}$.
}
\end{table}
%
%
\begin{table}
\begin{center}
\begin{tabular}{lllllll}
$\beta$   & & $a m_{ud}$     &  \multicolumn{2}{c}{ NP.(MeV) }         &  \multicolumn{2}{c}{ P.(MeV) } \\
          & & $( 10^{-4} )$  &  $p=1.8 {\rm GeV}$ & $p=2.6 {\rm GeV}$  &  $q^*=1/a$ &  $q^*=\pi/a$      \\
\hline
6.0       & & 9.72(40)              & 4.93(23) & 4.52(21)  & 3.37(15)   & 3.20(15)           \\
6.2       & & 6.86(45)              & 4.56(34) & 4.39(32)  & 3.36(25)   & 3.23(24)           \\
6.4       & & 5.38(23)              & 4.39(21) & 4.31(21)  & 3.45(16)   & 3.32(16)           \\
\multicolumn{3}{l}{$a^2\to 0$}     & 4.17(30) & 4.23(29)  & 3.46(23)   & 3.36(22)           \\ 
\end{tabular} 
\end{center}
\caption{ \label{Pesult_mud}
Final results for 
$m^{\overline{\rm MS}}_{ud}(2{\rm GeV})$
obtained with the non-perturbative (NP) and the perturbative (P) renormalization factors.  
The lattice bare quark mass $a m_{ud}$ is also listed.
}
\end{table}
%
%
%
%
\begin{table}
\begin{center}
\begin{tabular}{lllllll}
$\beta$     & &  $a m_s$       &  \multicolumn{2}{c}{ NP.(MeV) }        &  \multicolumn{2}{c}{ P.(MeV) }  \\
            & &  $( 10^{-2} )$ & $p=1.8 {\rm GeV}$ & $p=2.6 {\rm GeV}$  &  $q^*=1/a$  &  $q^*=\pi/a$      \\
\hline
\multicolumn{4}{l}{$K$ input} \\
6.0         & &  2.44(10)          & 123.8(5.7)   & 113.6(5.3)  & 84.6(3.9)   & 80.4(3.7)  \\  
6.2         & &  1.72(11)          & 114.4(3.3)   & 110.1(8.0)  & 84.5(6.2)   & 81.0(5.9)  \\
6.4         & &  1.350(57)         & 110.3(5.2)   & 108.2(5.1)  & 86.5(41)    & 83.4(3.9)  \\
\multicolumn{3}{l}{$a^2\to 0$}   & 104.7(7.4)   & 106.0(7.1)  & 86.9(5.6)   & 84.2(5.4)  \\
\hline
\multicolumn{4}{l}{$\phi$ input} \\
6.0         & & 2.73(15)          & 138.5(8.2)   & 127.1(7.5)  & 94.6(5.4)    & 90.0(5.2)  \\
6.2         & & 2.04(23)          & 136(16)      & 131(15)     & 100(12)      & 96(11)     \\
6.4         & & 1.597(97)         & 130.4(8.4)   & 128(8.3)    & 102.3(6.6)   & 98.7(6.3)  \\
\multicolumn{3}{l}{$a^2\to 0$}   & 128(11)      & 129(12)     & 105.5(9.1)   & 102.2(8.7) \\
\end{tabular} 
\end{center}
\caption{ \label{Pesult_ms}
The final results for the strage quark mass $m^{\overline{\rm MS}}_{s}(2{\rm GeV})$
obtained with $K$ meson or $\phi$ meson mass
to fix the bare strange quark masses $a m_s$.
}
\end{table}
%
%
%
%

\begin{thebibliography}{9}
%
%
\bibitem{review}
R. Gupta and T. Bhattacharya, Nucl. Phys. {\bf B}(Proc. Suppl.){\bf 63}, 95 (1998);
R.D. Kenway, hep-lat/9810054.
%
\bibitem{martinelli}
G. Martinelli {\it et.al.}, Nucl. Phys. {\bf B445}, 81 (1995).
%
\bibitem{jansen} 
K. Jansen {\it et.al.}, Nucl. Phys. {\bf B372}, 275 (1996); 
S. Capitani, hep-lat/9810063. 
%
\bibitem{gimenez} 
V. Gimen\'ez {\it et.al.}, hep-lat/9801028.
%
\bibitem{becirevic}
D. Becirevic {\it et.al.}, hep-lat/9807046.
%
\bibitem{perturbation} 
M. Golterman and J. Smit, Phys. Lett. {\bf 140B}, 392 (1984);
A. Patel and S. Sharpe, Nucl. Phys. {\bf B395}, 701 (1993);
N. Ishizuka and Y.Shizawa, Phys. Rev. {\bf D49}, 3519 (1994).
%
\bibitem{KS}
D. Daniel and S.N. Sheard, Nucl. Phys. {\bf B302}, 471 (1988).
%
\bibitem{ishizuka98} 
JLQCD Collaboration, S. Aoki {\it et.al. }, hep-lat/9809124.
%
\bibitem{kilcup97} For an earlier discussion, see, 
G. Kilcup, R. Gupta and S. Sharpe, Phys. Rev. {\bf D57}, 1654 (1998).
%
\bibitem{JLQCD_M}
JLQCD Collaboration, S. Aoki {\it et.al. }, Nucl. Phys. {\bf B}(Proc. Suppl.) {\bf 53}, 209 (1997).
%
\bibitem{lepage}
G.P. Lepage and P.B. Mackenzie, Phys. Rev. {\bf D 48}, 2250 (1993).
%
\bibitem{three_loop}
E. Franco and V. Lubicz, Nucl. Phys. {\bf B531}, 641 (1998);
T. van Ritbergen {\it et.al.},  Phys. Lett. {\bf B400}, 379 (1997);
J.A,M, Vermaseren {\it et.al.}, Phys. Lett. {\bf B405}, 327 (1997);
K.G. Chetyrkin, Phys. Lett. {\bf B404} 161 (1997).
%
\bibitem{CP-PACS}
CP-PACS Collaboration, hep-lat/9809146.
%
%
\end{thebibliography}
\end{document}